\documentclass[showpacs,preprint,amsmath,amssymb]{revtex4}
\usepackage{amsmath}
\usepackage{graphicx}
\usepackage{bm}
\usepackage{amsmath}
\usepackage{amsfonts}
\usepackage{amsbsy}
\usepackage{amssymb}
\usepackage{epsfig}

\begin{document}

\title{Dissipation can enhance quantum effects}
\author{Joachim Ankerhold}
\affiliation{Physikalisches Institut,
Albert-Ludwigs-Universit{\"a}t, D-79104 Freiburg, Germany }
\author{Eli Pollak}
\affiliation{Chemical Physics Department, Weizmann
Institute of Science, 76100, Rehovoth, Israel}
\date{\today}

\begin{abstract}
Usually one finds that dissipation tends to make a quantum system
more classical in nature. In this paper we study the effect of
momentum dissipation on a quantum system. The momentum of the
particle is coupled bilinearly to the momenta of a harmonic
oscillator heat bath. For a harmonic oscillator system we find that
the position and momentum variances for momentum coupling are
respectively identical to momentum and position variances for
spatial friction. This implies that momentum coupling leads to an
increase in the fluctuations in position as the temperature is
lowered, exactly the opposite of the classical like localization of
the oscillator, found with spatial friction. For a parabolic
barrier, momentum coupling causes an {\it increase} in the unstable
normal mode barrier frequency, as compared to the lowering of the
barrier frequency in the presence of purely spatial coupling. This
increase in the frequency leads to an enhancement of the thermal
tunneling flux, which below the crossover temperature becomes
exponentially large. The crossover temperature between tunneling and
thermal activation {\it increases} with momentum friction so that
quantum effects in the escape are relevant at higher temperatures.
\end{abstract}

\maketitle

\section{Introduction}

The effect of spatial dissipation on the classical
\cite{risken84} and quantum \cite{weiss95} dynamics of a
system is well understood. On a microscopic level,
dissipation arises from bilinear coupling of the system
coordinate to the displacement coordinates of a harmonic
bath. Classically the bath modes obey forced oscillator
equations of motion, which may be solved formally in terms
of the motion of the system. These are then inserted in
the system equation of motion which then takes the form of
a generalized Langevin equation.

Over twenty years ago, Caldeira and Leggett
\cite{caldeira83} took advantage of this equivalence to
study the effect of a dissipative bath on the quantum
dynamics of the system, paying special attention to the
quantum tunneling effect. Their central conclusion was that
dissipation reduces the tunneling probability, however, it
does not destroy it completely. Hence, the possibility of
observing macroscopic quantum tunneling.

The detrimental effect of an interaction between the
system and its environment on quantum phenomena makes
intuitive sense. Consider first the localization of a
particle in space. It is well known \cite{grabert88} that
the position variance of a dissipative harmonic oscillator
becomes smaller as the dissipation strength is increased.
In the limit of very large Ohmic friction, the bath can
localize the particle completely, without violating the
uncertainty principle \cite{ankerhold01}. The bath may be
thought of as creating an effective particle with a very
large mass, and such a heavy particle may be localized.

One has a similar picture of how the environment destroys
tunneling. For a dissipative parabolic barrier, it is well
known that diagonalization of the system-bath Hamiltonian
leads to an unstable mode, whose frequency decreases as
the coupling strength increases \cite{pollak86,pollak96}.
Tunneling occurs by transmission through this collective
unstable mode. Since its frequency is smaller, it is a
broader barrier, the action needed to cross it increases
and the tunneling probability decreases \cite{pollak86b}.
The same qualitative picture holds at low temperatures
below the crossover temperature separating the tunneling
and activated barrier crossing regimes
\cite{caldeira83,weiss95,pollak86c}. Spatial dissipation
also reduces the crossover temperature \cite{grabert84},
it is proportional to the collective mode barrier
frequency. This lowering also fits in with the general
observation that dissipation causes quantum systems to
behave more like classical systems \cite{caldeira83}.

All of these conclusions are based on the extensive study of the
quantum dynamics of dissipative systems, where the Hamiltonian can
be brought to the form of bilinear coupling between the system and
bath coordinates. There is a qualitative difference between this
type of spatial dissipative coupling and bilinear momentum coupling
of a system coupled to a bath \cite{cuccoli01}. Recently,
Makhnovskii and Pollak \cite{makhnovskii06} have shown that bilinear
momentum coupling leads to stochastic acceleration
\cite{fermi49,sturrock66} without any violation of the second law of
thermodynamics \cite{cole95}. In contrast to spatial coupling which
effectively increases the mass of the system, momentum coupling
reduces it and in the limit of an "Ohmic" coupling, the effective
mass goes to zero. Hence, the system can undergo stochastic
acceleration. This observation indicates that perhaps momentum
coupling can lead to some rather anti-intuitive quantum mechanical
results. If it reduces the effective mass, it should amplify quantum
effects rather than destroy them. This is the topic of this paper.

In Section II we study the classical and quantum dynamics of a
harmonic oscillator bilinearly coupled to the momentum of a harmonic
bath. We find that as already noted in a different context by
Cuccoli et al \cite{cuccoli01}, here too the effective mass is
reduced such that increasing the momentum coupling increases the
thermal variance of the position of the quantum particle instead of
decreasing it. In Section III we study the dynamics of a parabolic
barrier. Momentum coupling increases the thermal flux of the
particle across the barrier as compared to the thermal flux in the
absence of coupling. Most interestingly, the bath increases the
magnitude of the normal mode parabolic barrier frequency, implying
that it becomes thinner and therefore the tunneling flux through the
barrier increases. We show that this is indeed the case both above
and below the crossover temperature, which now increases with
increasing coupling strength. We end  in Section IV with a Summary.

\section{Momentum coupling and the harmonic
oscillator}\label{sec:2}

\subsection{Preliminaries}\label{sec:2a}

Our model is that of a harmonic oscillator with mass weighted
momentum $P$, coordinate $Q$ and harmonic frequency $\Omega $
interacting bilinearly with a harmonic oscillator heat bath through
the momentum \cite{makhnovskii06}. The Hamiltonian then takes the
form:
\begin{equation}
H=\frac{1}{2}\left[ P^{2}+\Omega ^{2}Q^{2}+\sum_{j=1}^{N}\left(
p_{j}-d_{j}P\right) ^{2}+\sum_{j=1}^{N}\omega _{j}^{2}x_{j}{}^{2}\right] \,,
\label{2.1}
\end{equation}%
where $p_{j},x_{j},\ \ j=1,...,N$ are the mass weighted momentum and
coordinate of the $j$-th bath oscillator whose frequency is $\omega
_{j}$. The $d_{j}$'s are the bilinear coupling coefficients to the
particle's momentum.

Before considering the dynamics of this Hamiltonian, it is
appropriate to put it into the context of previous studies of
dissipative systems. The coupling of the system to the bath through
the momentum of the bath has been studied previously in a variety of
contexts. In an early paper, Leggett \cite{leggett84} considered the
possibility of two coupling terms, taking the form
$Q\sum_{j=1}^Nd_jp_j+P\sum_{j=1}^Nc_jx_j$. He then distinguishes
between {\it normal} dissipation, where the Langevin equation of
motion is derived for the spatial system coordinate $Q$ and {\it
anomalous} dissipation, where the Langevin equation of motion is
derived for the spatial system momentum. Neither of these describe
the model Hamiltonian given in Eq. \ref{2.1} above. In our model,
the coupling term has the form $ P\sum_{j=1}^Nd_jp_j$, leading to
qualitatively different dynamics.

A different model has been considered by Ford et al \cite{ford88}.
In their case the coupling to the bath takes the quadratic form
$\left(P-\sum_{j=1}^Nd_jp_j\right)^2$. It describes the physics of
black body radiation in which the momentum of the particle is
coupled to the magnetic field of the vacuum radiation. Such coupling
also differs from that given in Eq. \ref{2.1}. The counter term
appearing in the model of Ford et al causes a coupling between the
bath modes themselves and allows by a change of variables
\cite{ford88} to recast the problem into one which is equivalent in
form to the standard dissipative Hamiltonian studied in detail in
Refs. \cite{caldeira83,grabert88}.

More recently, Cuccoli et al \cite{cuccoli01} have studied a
momentum coupling model which is identical to Eq. \ref{2.1}. They
term this model as {\it anomalous} dissipative coupling. However, as
already discussed above, the dissipation of this model differs from
the one studied by Leggett \cite{leggett84}. To distinguish between
the two, we have used the terminology {\it momentum dissipation} for
Eq. \ref{2.1}.

The formal solution of Hamilton's equations of motion for the $j$
-th bath oscillator is \cite{makhnovskii06}%
\begin{equation}
x_{j}(t)=x_{j}(0)\cos (\omega _{j}t)+\frac{\dot{x}_{j}(0)}{\omega _{j}}\sin
(\omega _{j}t)-\frac{d_{j}}{\omega _{j}}\int_{0}^{t}dt^{\prime }\dot{P}%
(t^{\prime })\sin [\omega _{j}(t-t^{\prime })]\,.  \label{2.2}
\end{equation}%
The equations of motion for the particle are
\begin{eqnarray}
\dot{Q} &=&P-\sum_{j=1}^{N}d_{j}\dot{x}_{j}  \label{2.3} \\
\dot{P} &=&-\Omega ^{2}Q  \label{2.4}
\end{eqnarray}%
Eq. \ref{2.4} together with:%
\begin{equation}
M\ddot{Q}(t)+\Omega ^{2}Q=Mf_{P}(t)\;-\int_{0}^{t}dt^{\prime }\dot{P}%
(t^{\prime })M\varphi _{P}(t-t^{\prime })  \label{2.5}
\end{equation}%
provide a generalized Langevin equation description for the motion
of a particle with the effective mass
\begin{equation}
M=(1+\sum_{j=1}^{N}d_{j}^{2})^{-1}\,.  \label{2.6}
\end{equation}%
The noise is represented by the momentum $f_{P}$ random acceleration:
\begin{equation}
f_{P}(t)=\sum_{j=1}^{N}d_{j}\omega _{j}^{2}\left[ x_{j}(0)\cos (\omega
_{j}t)+\frac{\dot{x}_{j}(0)}{\omega _{j}}\sin (\omega _{j}t)\right] \,.
\label{2.7}
\end{equation}%
which has zero mean. Its correlation function is:
\begin{equation}
\beta \left\langle f_{P}(t)f_{P}(0)\right\rangle
=\sum_{j=1}^{N}d_{j}^{2}\omega _{j}^{2}\cos \omega _{j}t\equiv \eta
_{P}(t)\,.  \label{2.8}
\end{equation}%
The brackets denote averaging with respect to the thermal distribution ($%
e^{-\beta H}$). Finally, in Eq.~(\ref{2.5}) we also used the
notation:
\begin{equation}
\ \ \ \varphi _{P}(t)=\int_{0}^{t}dt^{\prime }\eta _{P}(t^{\prime
})=\sum_{j=1}^{N}d_{j}^{2}\omega _{j}\sin \left( \omega _{j}t\right) \,.
\label{2.9}
\end{equation}

The solution of the generalized Langevin equation (\ref{2.5}) may be
obtained by means of Laplace transformation, $\hat{f}(s)=\int_{0}^{\infty
}dte^{-st}f(t)$. Using the relation $\hat{\varphi}_{P}(s)={\hat{\eta}_{P}(s)}%
/{s}$ one finds from Eq. (\ref{2.5}) that the Laplace transform of the
particle's coordinate is
\begin{equation}
{\hat{Q}}(s)=\frac{\dot{Q}(0)+sQ(0)+\hat{f}_{P}(s)}{s^{2}+\Omega ^{2}\left[
\frac{1}{M}-\frac{\hat{\eta}_{P}(s)}{s}\right] }=\frac{\dot{Q}(0)+sQ(0)+\hat{%
f}_{P}(s)}{s^{2}\left( 1+\Omega ^{2}\sum_{j=1}^{N}\frac{d_{j}^{2}}{%
s^{2}+\omega _{j}^{2}}\right) }\,.  \label{2.10}
\end{equation}%
Noting that
\begin{equation}
\left\langle \dot{Q}^{2}\right\rangle =\frac{\left\langle
P^{2}\right\rangle }{M}=\frac{1}{\beta M}  \label{2.11}
\end{equation}%
one readily finds that the classical velocity correlation function
is:
\begin{equation}
\left\langle \dot{Q}(t)\dot{Q}(0)\right\rangle ={\rm i.l.t.}\left(
\frac{s}{\beta}\cdot\frac{\frac{1}{M}-\frac{{\hat\eta}(s)}{s}}
{s^{2}+\Omega ^{2}\left[ \frac{1}{M}-\frac{\hat{\eta}%
_{P}(s)}{s}\right] }\right) .  \label{2.12}
\end{equation}
where i.l.t. stands for "inverse Laplace transform".

To express results in the continuum limit it is useful to define a momentum
spectral density as%
\begin{equation}
J_{P}(\omega )=\frac{\pi }{2}\sum_{j=1}^{N}d_{j}^{2}\omega _{j}^{3}\delta
(\omega -\omega _{j})  \label{2.13}
\end{equation}%
where $\delta (x)$ is the Dirac "$\delta $" function. As a result, the
momentum function $\eta _{P}(t)$ may be expressed in terms of the spectral
density as:%
\begin{equation}
\eta _{P}(t)=\frac{2}{\pi }\int_{0}^{\infty }d\omega \frac{J_{P}(\omega )}{%
\omega }\cos \left( \omega t\right) .  \label{2.14}
\end{equation}

\subsection{The normal mode transformation}\label{sec:2b}

Additional insight as well as solution of the associated
quantum dynamics is facilitated by considering the normal
modes representation. The Hamiltonian given in
Eq.~(\ref{2.1}) has a quadratic form and so may be
diagonalized. For
this purpose we define frequency weighted coordinates and momenta as%
\begin{eqnarray}
\bar{Q} &=&\Omega Q,\text{ \ \ }\bar{P}=P/\Omega  \label{2.15} \\
\bar{x}_{j} &=&\omega _{j}x_{j},\text{ \ \ }\bar{p}_{j}=p_{j}/\omega _{j},%
\text{ \ \ }j=1,...,N  \label{2.16}
\end{eqnarray}%
so that the coordinate part of the Hamiltonian has unit frequency:%
\begin{eqnarray}
H=\frac{1}{2}\left[ \Omega ^{2}\bar{P}^{2}+\sum_{j=1}^{N}\omega
_{j}^{2}\left( \bar{p}_{j}-\frac{d_{j}}{\omega _{j}}\Omega \bar{P}\right)
^{2}+\bar{Q}^{2}+\sum_{j=1}^{N}\bar{x}_{j}{}^{2}\right] .  \label{2.17}
\end{eqnarray}
The $N+1$ normal modes and associated momenta are denoted as $%
y_{j},p_{y_{j}};j=0,...,N$ such that the normal mode form of the Hamiltonian
is:
\begin{eqnarray}
H=\frac{1}{2}\sum_{j=0}^{N}\left( \lambda
_{j}^{2}p_{y_{j}}^{2}+y_{j}^{2}\right)  \label{2.18}
\end{eqnarray}
and the $\lambda _{j}$'s are the normal mode frequencies. This
transformation implies that the vector of normal mode momenta $\mathbf{p}%
_{y} $ is an orthogonal transformation of the frequency weighted momenta
such that
\begin{eqnarray}
\mathbf{p}_{y}=\mathbf{U}\left(
\begin{array}{c}
\bar{P} \\
\mathbf{\bar{p}}%
\end{array}
\right)  \label{2.19}
\end{eqnarray}
where $\mathbf{U}$ is an $(N+1)\times(N+1)$ orthogonal
transformation matrix.

Following the same considerations as in the appendix of
Ref.~\cite{levine88} one readily finds that the normal mode
frequencies are the $N+1$ solutions of the
equation:%
\begin{eqnarray}
\lambda _{k}^{2}=\frac{\Omega ^{2}}{1+\Omega ^{2}\sum_{j=1}^{N}\frac{%
d_{j}^{2}}{\omega _{j}^{2}-\lambda _{k}^{2}}}.  \label{2.20}
\end{eqnarray}
The elements of the transformation matrix are then given by:%
\begin{eqnarray}
u_{kj} &=&\frac{d_{j}\omega _{j}\Omega }{\omega _{j}^{2}-\lambda _{k}^{2}}%
u_{k0},\text{ \ \ }j=1,...,N,\text{ \ \ }k=0,...,N  \label{2.21} \\
u_{k0}^{2} &=&\left( 1+\Omega ^{2}\sum_{j=1}^{N}\frac{d_{j}^{2}\omega
_{j}^{2}}{\left( \omega _{j}^{2}-\lambda _{k}^{2}\right) ^{2}}\right) ^{-1},%
\text{ \ \ }k=0,...,N.  \label{2.22}
\end{eqnarray}%
By considering the $00$ element of the $(N+1)$x$(N+1)$ matrix $(\mathbf{T}%
^{\prime \prime }+s^{2}\mathbf{I)}^{-1}$, (where $\mathbf{T}^{\prime \prime
} $ is the matrix of second derivatives of the kinetic energy of the
Hamiltonian with respect to the frequency scaled momenta) one finds the important identity:%
\begin{eqnarray}
\sum_{j=0}^{N}\frac{u_{j0}^{2}}{s^{2}+\lambda _{j}^{2}}=\left[ s^{2}+\Omega
^{2}\left( \frac{1}{M}-\frac{\hat{\eta}_{P}(s)}{s}\right) \right] ^{-1}.
\label{2.23}
\end{eqnarray}
This identity then leads directly to all classical results of
interest in the continuum limit.

For this purpose we also define a normal mode momentum function as:%
\begin{eqnarray}
K(t)=\sum_{j=0}^{N}u_{j0}^{2}\cos (\lambda _{j}t).  \label{2.24}
\end{eqnarray}
A spectral density of the normal modes is then defined as \cite{rips90}:%
\begin{eqnarray}
\Upsilon (\lambda )=\frac{\pi }{2}\sum_{j=0}^{N}u_{j0}^{2}\lambda _{j}\left[
\delta \left( \lambda -\lambda _{j}\right) -\delta \left( \lambda +\lambda
_{j}\right) \right]  \label{2.25}
\end{eqnarray}
One now notes that the Laplace transform of the normal mode momentum
function may be expressed directly in terms of the original momentum
function:%
\begin{eqnarray}
\frac{\hat{K}(s)}{s}=\sum_{j=0}^{N}\frac{u_{j0}^{2}}{s^{2}+\lambda _{j}^{2}}=%
\left[ s^{2}+\Omega ^{2}\left( \frac{1}{M}-\frac{\hat{\eta}_{P}(s)}{s}%
\right) \right] ^{-1}.  \label{2.26}
\end{eqnarray}
Using the Fourier decomposition of the Dirac $\delta $ function ($\pi \delta
(\lambda )=\int_{0}^{\infty }dt\cos (\lambda t)$) we find that the spectral
density of the normal modes may also be expressed in the continuum limit as:
\begin{eqnarray}
\Upsilon (\lambda )=\mathrm{{Re}\left[ \lambda \hat{K}(i\lambda )\right] .}
\label{2.27}
\end{eqnarray}
We further note that at equilibrium:
\begin{eqnarray}
\left\langle y_{j}^{2}\right\rangle &=&k_{B}T,\text{\ \ \ \ \ \ \ \
\
\ \ \ \ \ \ \ \ \ \ \ }j=0,...,N  \label{2.28} \\
\text{\ }\left\langle \dot{y}_{j}^{2}\right\rangle &=&\lambda
_{j}^{4}\left\langle p_{y_{j}}^{2}\right\rangle =\lambda _{j}^{2}k_{B}T,%
\text{ \ \ }j=0,...,N\text{\ .}  \label{2.29}
\end{eqnarray}

With these preliminaries it becomes straightforward to solve for thermal
correlation functions. Since the Hamiltonian is diagonal in the normal modes
one has trivially that the solution of the j-th normal mode is:%
\begin{eqnarray}
y_{j}(t)=y_{j}(0)\cos (\lambda _{j}t)+\frac{\dot{y}_{j}(0)}{\lambda _{j}}%
\sin (\lambda _{j}t),\text{ \ \ \ }j=0,...,N.  \label{2.30}
\end{eqnarray}
The system coordinate is just a linear combination of the normal modes%
\begin{eqnarray}
Q(t)=\frac{\bar{Q}(t)}{\Omega }=\frac{\sum_{j=0}^{N}u_{j0}y_{j}(t)}{\Omega }
\label{2.31}
\end{eqnarray}
so that
\begin{eqnarray}
\left\langle Q(t)Q(0)\right\rangle =\frac{\sum_{j=0}^{N}u_{j0}^{2}\left%
\langle y_{j}^{2}\right\rangle \cos (\lambda _{j}t)}{\Omega ^{2}}=\frac{K(t)%
}{\beta \Omega ^{2}}.  \label{2.32}
\end{eqnarray}
Similarly%
\begin{eqnarray}
\left\langle \dot{Q}(t)\dot{Q}(0)\right\rangle =-\frac{\ddot{K}(t)}{\beta
\Omega ^{2}}.  \label{2.33}
\end{eqnarray}
These properties of the normal mode transformation become
very useful also when considering the barrier crossing
dynamics, as described in Section \ref{sec:3}, below.
\newline

\subsection{Quantum dynamics}\label{sec:2c}

In the quantum regime one may work with the normal modes and use the
quantum instead of the classical expressions (\ref{2.28}),
(\ref{2.29}). For the equilibrium variances the procedure is then as
follows. From the quantum mechanical expressions
\begin{equation}
\left\langle y_{j}^{2}\right\rangle =\frac{\hbar \lambda
_{j}}{2}\coth \left( \frac{\hbar \beta \lambda
_{j}}{2}\right)
\end{equation}%
it follows that
\begin{equation}
\left\langle Q^{2}\right\rangle =\frac{\hbar }{2\Omega ^{2}}%
\sum_{j=0}^{N}u_{j0}^{2}\lambda _{j}\coth \left(
\frac{\hbar \beta \lambda _{j}}{2}\right) .
\end{equation}%
Using the decomposition \cite{rhyzhik}
\begin{equation}
\coth (\pi x)=\frac{1}{\pi x}+\frac{2x}{\pi }\sum_{k=1}^{\infty }\frac{1}{%
x^{2}+k^{2}}
\end{equation}%
 we arrive at
\begin{equation}
\left\langle Q^{2}\right\rangle =\frac{1}{\beta \Omega
^{2}}\left[ 1+2\sum_{j=0}^{N}u_{j0}^{2}\lambda
_{j}^{2}\sum_{k=1}^{\infty }\frac{1}{\lambda _{j}^{2}+\nu
_{k}{}^{2}}\right]
\end{equation}%
with the Matsubara frequencies $\nu _{k}=2\pi k/(\hbar
\beta )$. Interchanging the sums, using the fact
$\sum_{j=0}^{N}u_{j0}^{2}=1$,  and the identity
(\ref{2.23}) gives us
\begin{equation}
\left\langle Q^{2}\right\rangle
=\frac{1}{\beta \Omega ^{2}}\left[ 1+2\sum_{k=1}^{\infty }\left( \frac{%
\Omega ^{2}\left( \frac{1}{M}-\frac{\hat{\eta}_{P}(\nu _{k})}{\nu _{k}}%
\right) }{\nu _{k}^{2}+\Omega ^{2}\left( \frac{1}{M}-\frac{\hat{\eta}%
_{P}(\nu _{k})}{\nu _{k}}\right) }\right) \right]\, .\label{2.38}
\end{equation}%
It is worthwhile to derive a similar expression for the momentum. In this case:%
\begin{equation}
\left\langle P^{2}\right\rangle =\Omega
^{2}\sum_{j=0}^{N}u_{j0}^{2}\left\langle
p_{y_{j}}^{2}\right\rangle
\end{equation}%
with $\langle p_{y_{j}}^{2}\rangle =(\hbar/2\lambda
_{j})\coth(\hbar \beta \lambda_{j}/2) $. It follows that
\begin{eqnarray}
\left\langle P^{2}\right\rangle  &=&\frac{\hbar }{2}\Omega ^{2}\sum_{j=0}^{N}%
\frac{u_{j0}^{2}}{\lambda _{j}}\coth \left( \frac{\hbar \beta \lambda _{j}}{2%
}\right) \nonumber\\
&=&\frac{1}{\beta }\left[ 1+2\Omega ^{2}\sum_{j=0}^{N}\sum_{k=1}^{\infty }%
\frac{u_{j0}^{2}}{\lambda _{j}^{2}+\nu
_{k}{}^{2}}\right]\, ,
\end{eqnarray}%
where we used the identity (obtained from Eq. \ref{2.23} in the
limit that $s\rightarrow 0$) that
\begin{equation}
\sum_{j=0}^{N}\frac{u_{j0}^{2}}{\lambda
_{j}^{2}}=\frac{1}{\Omega ^{2}}.
\end{equation}%
Finally using again identity (\ref{2.23}) one obtains
\begin{equation}
\left\langle P^{2}\right\rangle =\frac{1}{\beta }\left[
1+2\Omega
^{2}\sum_{k=1}^{\infty }\frac{1}{\nu _{k}{}^{2}+\Omega ^{2}\left( \frac{1}{M}%
-\frac{\hat{\eta}_{P}(\nu _{k})}{\nu _{k}}\right) }\right]\,
.\label{2.42}
\end{equation}%

It is instructive to derive the quantum mechanical correlations
along an alternative route, which for spatial friction has been
discussed in Ref. \cite{weiss95,grabert84} and only exploits
fundamental principles of quantum statistical mechanics. Namely,
since the equations of motion  for the Heisenberg operators $Q(t)$,
$P(t)$ are linear, the following is true: (i) Due to Ehrenfest's
theorem the quantum mechanical averages obey classical equations of
motion, (ii) correlation functions can be obtained from the quantum
version of the fluctuation dissipation theorem, and (iii) mean
values and second order correlations completely determine the
quantum dynamics since all random forces are related to a stationary
Gaussian process.

When using (ii) one has to take into account that according to
(\ref{2.3}) and in contrast to spatial friction the time derivative
of $ Q$ is not identical to $P$. Hence, to apply the fluctuation
dissipation theorem we do not start from the equation of motion in
position (\ref{2.5}), but from the corresponding expression in
momentum, i.e.,
\begin{equation}
\langle\ddot{P}(t)\rangle+\frac{\Omega ^{2}}{M}\, \langle
P(t)\rangle-\Omega^2\int_{0}^{t}dt^{\prime }
\langle{P}(t^{\prime })\rangle \varphi _{P}(t-t^{\prime
})=0 \label{plangevin}
\end{equation}%
and calculate according to (i) the classical response
\begin{equation}
\langle P(t)\rangle=\int_0^t dt' \ \chi_P(t-t')\, F(t')\, ,
\label{response}
\end{equation}
to an external force $F(t)$ applied for $t>0$. In Fourier
space the above equation reads $\langle
\tilde{P}(\omega)\rangle=\tilde{\chi}_P(\omega)\,
\tilde{F}(\omega)$ so that
\begin{equation}
\tilde{\chi}_P(\omega)=\frac{\Omega^2}{\Omega^2/M-\omega^2-\Omega^2
\tilde{\varphi}_P(\omega)}\label{moresp}
\end{equation}
with
$\tilde{\varphi}_P(\omega)=\hat{\eta}_P(-i\omega)/(-i\omega)$.

According to (ii)  it is now the symmetrized momentum
correlation $S_P(t)=(1/2) \langle P(t)P(0)+P(0)
P(t)\rangle$, which is related to the imaginary part of
the response function $\tilde{\chi}_P=\tilde{\chi}_P'+i
\tilde{\chi}_P''$ via
\begin{equation}
\tilde{S}_P(\omega)=\hbar \coth(\omega\hbar\beta/2)\,
\tilde{\chi}_P''(\omega)\, .\label{dissfluc}
\end{equation}
Further, due to (\ref{response}) the anti-symmetrized momentum
correlation function $A_P(t)=(1/i) [P(t), P(0)]$ is related to the
response function via
\begin{equation}
\chi_P(t)=-\frac{2}{\hbar}\,\theta(t)\, A_P(t)
\end{equation}
with the step function $\theta(\cdot)$. Thus,  in the time
domain we arrive at the general expressions
\begin{equation}
S_P(t)=\frac{\hbar}{2\pi}\int_{-\infty}^\infty d\omega\
\tilde{\chi}_P''(\omega)\, {\rm coth}(\omega\hbar\beta/2)\,
\cos(\omega t) \label{st}
\end{equation}
and
\begin{equation}
A_P(t)=-\frac{\hbar}{2\pi}\int_{-\infty}^\infty d\omega\
\tilde{\chi}_P''(\omega)\, \sin(\omega t) \, ,\label{at}
\end{equation}
from which following (iii) all real-time correlations can
be derived, e.g.\ $\langle Q(t) Q(0)\rangle= \langle
\dot{P}(t) \dot{P}(0)\rangle/\Omega^4=[-\ddot{S}_P(t)-i
\ddot{A}_P(t)]/\Omega^4$. For analytical calculations it is
sometimes more convenient to work with representations
based on Laplace transforms, namely,
\begin{eqnarray}
\hat{A}_P(s)&=&-\frac{\hbar}{2} \ \hat{\chi}_P(s)\nonumber\\
\hat{S}_P(s)&=&\frac{1}{\beta}\sum_{n=-\infty}^\infty
\frac{s}{\nu_n^2-s^2}\left[\hat{\chi}_P(s)-\hat{\chi}_P(|\nu_n|)\right]\,
,
 \label{laplaceas}
\end{eqnarray}
where we used that $\hat{\chi}_P(s)=\tilde{\chi}_P(i s)$.

By considering $S_P(0)=-\lim_{s\to \infty} s \hat{S}_P(s)$
one obtains the equilibrium variance in momentum
\begin{equation}
\langle P^2\rangle_\beta=\frac{1}{\beta}
\sum_{n=-\infty}^\infty \, \hat{\chi}_P(
|\nu_n|)=\frac{1}{\beta}+\frac{2\Omega^2}{\beta}\sum_{n=1}^\infty
\,\frac{1}{\Omega^2
(\frac{1}{M}-\frac{\hat{\eta}(\nu_n)}{\nu_n})+\nu_n^2}\, .
\label{p2}
\end{equation}
which is of course, identical to Eq.~(\ref{2.42}). For the
position $\langle Q^2\rangle_\beta=-\ddot{S}_P(0)/\Omega^4$
 one must be careful when taking the limit in $\ddot{S}_P(0)=\lim_{s\to\infty} s^3 \hat{S}_P(s)$
due to singularities which must be properly subtracted. One
gets
\begin{eqnarray}
\langle Q^2\rangle_\beta
&=&\frac{2}{\beta\Omega^4}\sum_{n=-\infty}^\infty
\left[\Omega^2-\nu_n^2 \hat{\chi}_P(|\nu_n|)\right]\nonumber\\
&=&\frac{1}{\Omega^2\beta}+\frac{2}{\beta}\sum_{n=1}^\infty
\frac{\frac{1}{M}-\frac{\hat{\eta}_P(\nu_n)}{\nu_n}}
{\nu_n^2+\Omega^2\left[\frac{1}{M}-\frac{\hat{\eta}_P(\nu_n)}{\nu_n}\right]}
\, . \label{q2}
\end{eqnarray}
which is identical to Eq.~(\ref{2.38}). When comparing the
above expressions with those derived for spatial friction
\cite{grabert87,weiss95}, one observes that they can be
related to each other by $\Omega Q\leftrightarrow
 P$, which, of course, is a direct consequence of
the symmetry of the Hamiltonian (\ref{2.1}). This means
though, that all findings known for spatial friction and
e.g.\ ohmic damping can be directly translated to momentum
friction. We will discuss explicit results in the next
Section.

\subsection{An example}\label{sec:2d}

To obtain a feeling for the various results presented in this
Section it is worthwhile to consider the specific case of a momentum
density with a cutoff:%
\begin{equation}
J_{P}(\omega )=\frac{\pi }{2}\gamma \omega ^{3}\theta
\left( \omega _{c}-\omega \right) \label{spectral}
\end{equation}
where $\theta (x)$ is the Heaviside function. One then finds that the
Laplace transform of the momentum function is%
\[
\hat{\eta}_{P}(s)=\gamma s\left[ \omega _{c}-s\arctan
\left( \frac{\omega _{c}}{s}\right)\right]
\]%
and the mass factor
\[
M=\left( 1+\gamma \omega _{c}\right) ^{-1}
\]%
explicitly demonstrating that in the limit that the cutoff frequency
goes to infinity, the effective mass goes to zero.

Now, let us first look at the friction dependence of the
Laplace transform $\hat{\chi}_P(s)$ which directly provides
the momentum variance and the functions $A_P(t)$ and
$S_P(t)$. From
\begin{equation}
\frac{1}{M}-\frac{\hat{\eta}_P(s)}{s}=1+\gamma s\, {\rm
arctan}(\omega_c/s)\,
\end{equation}
we see  that in the limit $\omega_c\to 0$ the friction term
in $\hat{\chi}_P$ combines with the potential term to yield
the effective frequency $\Omega_{\rm eff}=\Omega
\sqrt{1+\gamma \omega_c}$ of an undamped harmonic
oscillator. The corresponding variance $\langle P^2\rangle$
as well as $A_P(t), S_P(t)$ are thus trivial. In the
opposite limit of very large $\omega_c\gg \Omega,
\gamma\Omega^2$ one has in $\hat{\chi}_P$ the expression
$s^2+\Omega^2+\Omega^2\gamma \pi s/2$, which coincides
with the form of the response function for spatial
friction in the ohmic case (friction constant $\gamma_s$)
with the translation $\gamma_s=\Omega^2\gamma \pi/2$.
Thus, the corresponding results can be read off from the
literature \cite{grabert88,weiss95}. In particular, for
large cutoff and $\gamma\Omega^2\pi/2\geq 1$ the time
dependent correlations decay to zero monotonously, while
for $\gamma\Omega^2\pi/2<1$ they decay via damped
oscillations. In contrast to $A_P(t)$ which displays only
classical dynamics, $S_P(t)$ contains an additional
contribution depending on the Matsubara frequencies that
becomes relevant at lower temperatures. In the limit of
zero temperature $S_P(t)$ then decays in time no longer
exponentially but algebraically $\propto 1/t^2$. Further
in the limit of zero temperature one has for the momentum
variance
\begin{equation}
\langle
P^2\rangle_0\approx\frac{2\hbar}{\gamma\pi^2}\ln\left(\frac{\Omega^2\gamma^2\pi^2}{2}\right)\,
.  \label{p2zero}
\end{equation}
In essence, momentum friction gives for momentum correlations the
same results as spatial coupling for position correlations. The same
is true when the ratios $\omega_c/\Omega$, $\omega_c/\gamma\Omega^2$
decrease, but are still large, and one compares momentum friction
with $\gamma,\omega_c$ with the well-known Drude model for spatial
friction, $J_D=\gamma_s \omega \exp(-\omega/\omega_c)$, with
$\gamma_s=\gamma\Omega^2\pi/2$.

Let us now turn to $\langle Q^2\rangle$, for which the limit
$\omega_c\to \infty$ cannot be taken. We gain
\begin{equation}
\langle Q^2\rangle=\langle
P^2\rangle/\Omega^2+\frac{2\Omega^4\gamma}{\beta}\sum_{n=1}^\infty
\frac{\nu_n {\rm
arctan}(\omega_c/\nu_n)}{\nu_n^2+\Omega^2+\Omega^2\gamma\nu_n
{\rm arctan}(\omega_c/\nu_n)}\, , \label{q2exact}
\end{equation}
which for zero temperature, where the sum over Matsubara
frequencies must be replaced by integrals according to
$(2/\hbar\beta)\sum f(\nu_n)\to(1/\pi)\int d\nu f(\nu)$,
reads for strong friction (but still $\omega_c\gg
\gamma\Omega^2$)
\begin{equation}
\langle
Q^2\rangle_0\approx\frac{\hbar\Omega^4\gamma}{2}\ln\left(\frac{2\omega_c}{\pi
\gamma\Omega^2}\right) \, . \label{q2zero}
\end{equation}
When dealing with spatial friction, the momentum variance for a
Drude model with $\gamma_s=\gamma\Omega^2\pi/2$, is given by
$\langle P^2\rangle_{\rm spatial}\approx (\hbar
\Omega^2\gamma/2)\ln(\omega_c/\Omega)$ \cite{weiss95}. One thus
notes that the role of the momentum and position variances are
interchanged when considering the momentum coupling model.
\begin{figure}
\begin{center}
\vspace*{0.1cm} \epsfig{file=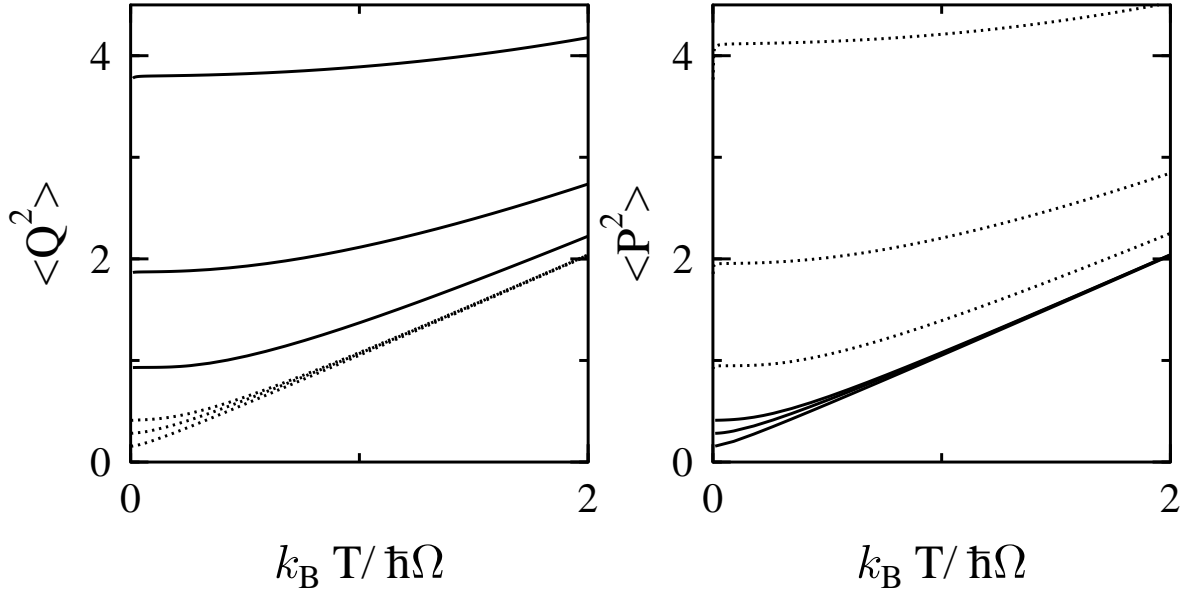, height=8cm}
\caption[]{\label{fignew1}Position variances (left, scaled
with $\hbar/\Omega$) and momentum variances (right, scaled
with $\hbar\Omega$) for momentum friction (solid) and for
the corresponding spatial case (dotted) for various values
of the friction strength $\gamma\Omega$=0.5, 2, 7: left,
solid from bottom to top; left dotted from top to bottom;
right, solid from top to bottom; right, dotted from bottom
to top. The cutoff frequency is $\omega_c/\Omega=10$. For
the spatial case a model with Drude damping (cutoff
$\omega_c$) is used with equivalent friction strength
$\gamma_s=\gamma\Omega^2\pi/2$.}
\end{center}
\end{figure}
In Fig.~\ref{fignew1} the position and momentum variances are shown
for momentum coupling with the spectral density (\ref{spectral})
together with spatial coupling with a Drude spectral density,
$J_D=\gamma_s\omega \exp(-\omega/\omega_c)$. Note that the latter
one is, apart from its widespread use, a sort of minimal model for
spatial friction leading to well-behaved variances. In case of
momentum coupling the spectral density (\ref{spectral}) plays a
similar role. Both models give identical results for sufficiently
large cutoff and $\gamma_s=\Omega^2\gamma \pi/2$, i.e.\
$\Omega^2\langle P^2\rangle_{\rm momentum}(\gamma)=\langle
Q^2\rangle_{\rm spatial}(\Omega^2\gamma \pi/2)$. They differ though
in the low temperature range for the variances $\langle
Q^2\rangle_{\rm momentum}$, $\langle P^2\rangle_{\rm spatial}$,
which are only well-behaved for finite cutoff. Of course, when for
spatial friction a spectral density is taken that produces the same
damping dependence in the response function as the one given by
(\ref{spectral}), we would obtain fully identical results.
Accordingly, position variances are enhanced for momentum coupling
and suppressed for spatial coupling and vice versa for the momentum
variances. One also observes that at low temperatures the momentum
variance for momentum friction is suppressed compared to the
position variance for spatial friction within a Drude model [see
(\ref{q2zero})]. Accordingly, the uncertainty product for $T=0$
\begin{equation}
\langle Q^2\rangle_0 \langle P^2\rangle_0 =
\frac{\hbar^2}{\pi^2}\ln\left(\frac{2\omega_c}{\pi\gamma\Omega^2}\right)\,
\ln\left(\frac{\pi^2\gamma^2\Omega^2}{2}\right)\,
\end{equation}
is smaller compared to the spatial case (Drude damping) by
the factor
$\left[1-\ln(\pi\gamma\Omega/2)/\ln(\omega_c/\Omega)\right]$.

\section{Classical and quantum rate theory}\label{sec:3}

\subsection{Classical rate theory in the presence of momentum coupling}\label{sec:3a}

We first consider the case of a parabolic barrier Hamiltonian:%
\begin{equation}
H=\frac{1}{2}\left[ P^{2}-\Omega ^{\ddag 2}Q^{2}+\sum_{j=1}^{N}\left(
p_{j}-d_{j}P\right) ^{2}+\sum_{j=1}^{N}\omega _{j}^{2}x_{j}{}^{2}\right] .
\label{3.1}
\end{equation}%
In conventional Transition State Theory (TST) one uses the system coordinate
as the reaction coordinate and the dividing surface is taken to be
perpendicular to it. The thermal unidirectional flux through the dividing
surface is \cite{keck67,pechukas76}:%
\begin{equation}
F_{TST}=\int_{-\infty }^{\infty
}dPdQ\prod\limits_{j=1}^{N}dp_{j}dx_{j}e^{-\beta H}\delta (Q)\dot{Q}\theta (%
\dot{Q})=\frac{1}{\beta \sqrt{M}}\prod\limits_{j=1}^{N}\left( \frac{2\pi }{%
\beta \omega _{j}}\right) .  \label{3.2}
\end{equation}%
It is noteworthy that the thermal flux through the dividing surface is
larger by the factor $1/\sqrt{M}$ as compared to the conventional TST flux
in the presence of only spatial coupling to the harmonic bath ($F_{0}=\sqrt{M%
}F_{TST}$). Momentum coupling causes an \textit{enhancement} of the reaction
rate.

The minimal unidirectional thermal flux is obtained by transforming the
Hamiltonian to normal modes. Due to the negative force constant associated
with the barrier, one will now find after diagonalization $N$ stable modes
with frequencies $\lambda _{j}$ and one unstable mode with barrier frequency
$\lambda ^{\ddag }$. The VTST flux is obtained by considering the flux
perpendicular to the unstable mode and one finds that:%
\begin{equation}
F_{VTST}=\frac{1}{\beta }\prod\limits_{j=1}^{N}\left( \frac{2\pi }{\beta
\lambda _{j}}\right) .  \label{3.3}
\end{equation}%
By considering the determinant of the second derivative matrix of the
kinetic energy in the normal mode representation and in the original
coordinates one readily finds the identity%
\begin{equation}
\lambda ^{\ddag 2}\prod\limits_{j=1}^{N}\lambda _{j}^{2}=\Omega ^{\ddag
2}\prod\limits_{j=1}^{N}\omega _{j}^{2}.  \label{3.4}
\end{equation}%
As a result the ratio of the VTST flux to the TST flux is%
\begin{equation}
\frac{F_{VTST}}{F_{TST}}=\sqrt{M}\frac{\lambda ^{\ddag }}{\Omega ^{\ddag }}.
\label{3.5}
\end{equation}

The analog of the Kramers-Grote-Hynes equation
\cite{kramers40,grote80} for the normal mode barrier frequency in
the presence of spatial diffusion is obtained from Eq.~(\ref{2.20}),
except that one substitutes the stable mode frequency with the
unstable mode frequency. After a bit of rearranging one finds:%
\begin{equation}
\lambda ^{\ddag 2}+\frac{\Omega ^{\ddag 2}\hat{\eta}_{P}(\lambda ^{\ddag })}{%
\lambda ^{\ddag }}=\frac{\Omega ^{\ddag 2}}{M}  \label{3.6}
\end{equation}%
from which it follows that $\sqrt{M}\frac{\lambda ^{\ddag }}{\Omega ^{\ddag }%
}\leq 1$ and as expected the VTST flux is lower than the TST flux. Note
however that Eq.~(\ref{3.6}) may also be rewritten as:%
\begin{equation}
\frac{\lambda ^{\ddag 2}}{\Omega ^{\ddag 2}}=1+\lambda ^{\ddag
2}\sum_{j=1}^{N}\frac{d_{j}^{2}}{\omega _{j}^{2}+\lambda ^{\ddag 2}}
\label{3.7}
\end{equation}%
showing that momentum coupling leads to a \textit{thinning} of the barrier,
instead of the usual broadening of the barrier found as a result of spatial
coupling. In fact the ratio $\frac{F_{VTST}}{F_{0}}=\frac{\lambda ^{\ddag }}{%
\Omega ^{\ddag }}$, implying that even the minimal VTST flux is \textit{%
larger }then the flux in the absence of coupling. This is proof that
classically, momentum coupling leads to an increase of the thermal
parabolic barrier crossing rate, as compared with the absence of
coupling.

\begin{figure}
\begin{center}
\vspace*{0.1cm} \epsfig{file=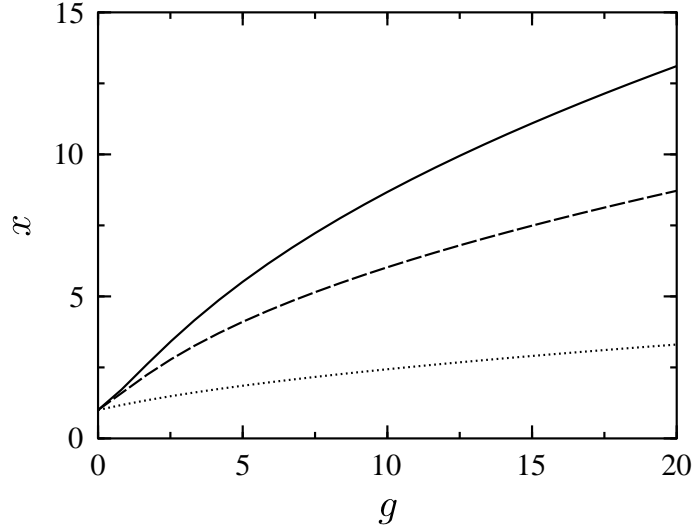, height=7cm}
\caption[]{\label{fig_1}Classical enhancement of the
reactive flux as a result of momentum coupling to a
harmonic bath: The parabolic barrier frequency is plotted
as a function of the coupling strength for three different
values of the cutoff frequency. The reduced barrier
frequency is also the ratio of the exact thermal
unidirectional flux to the flux in the absence of coupling
to the bath. The solid line is for $w_c=10$, the dashed
line for $w_c=4$ and the dotted line is for $w_c=0.5$.}
\end{center}
\end{figure}
\begin{figure}
\begin{center}
\vspace*{0.1cm} \epsfig{file=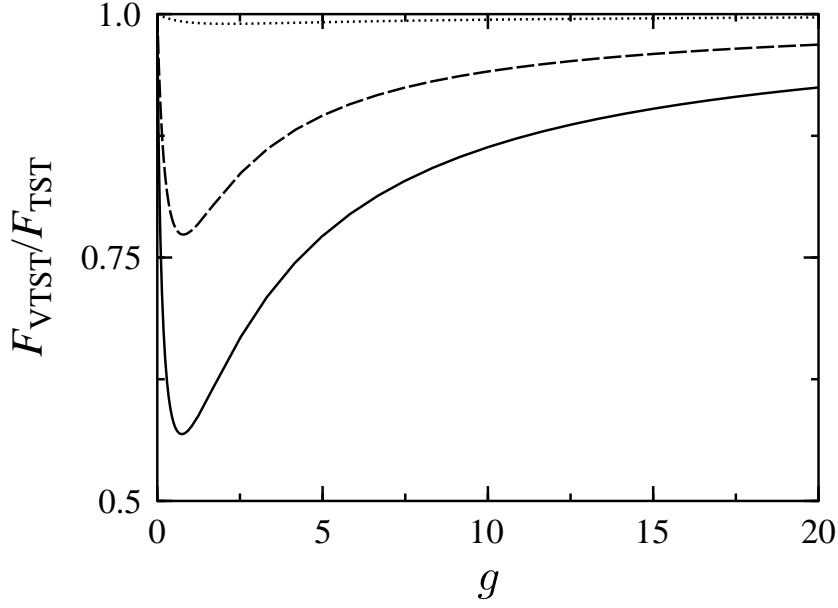, height=8cm}
\caption[]{\label{fig_2}VTST solution for the classical
transmission factor through a parabolic barrier potential
as compared to choosing the system coordinate as the
reaction coordinate. The three lines correspond to the
same values of the cutoff frequency as in
Fig.~\ref{fig_1}.}
\end{center}
\end{figure}

It is instructive to consider a specific example, namely
the spectral density given in Eq.~(\ref{spectral}). Using
the reduced values $x=$ $\lambda ^{\ddag }/\Omega ^{\ddag
},w_{c}=\omega _{c}/\Omega ^{\ddag }$ and $g=\gamma \Omega
^{\ddag }$ we plot in Fig.~\ref{fig_1} the reduced barrier
height as a function of the reduced momentum coupling
coefficient $g$ for three representative values of the
reduced cutoff frequency $w_{c}$. As noted from the
figure, only when the cutoff frequency is large does one
get an appreciable increase in the barrier frequency. In
the limit of $g\gg w_{c}$ one has that $x\sim
\sqrt{gw_{c}}$. In Fig.~\ref{fig_2} we then plot the ratio
$F_{VTST}/F_{TST}$ for the same parameter range as in
Fig.~\ref{fig_1}. One notes that there is an appreciable
effect on the transmission factor only when the cutoff
frequency is much larger than unity. In contrast to the
spatial coupling case, here the transmission factor tends
to unity both in the weak and strong momentum coupling
limits.

\subsection{Quantum rate theory in the presence of momentum coupling}\label{sec:3b}

Given the fact that momentum coupling causes a decrease in
the barrier width, it is interesting to study whether it
could then cause an increase in the tunneling probability,
since the tunneling probability is exponentially sensitive
to the width of the barrier. We will consider two cases.
The first is transmission through a parabolic barrier. The
second will be consideration of tunneling through an
anharmonic barrier in the limit of weak momentum coupling
and at temperatures which are below the crossover
temperature between tunneling and thermal activation. We
will see that the two limits give qualitatively identical
results. Above the crossover temperature, where the
parabolic barrier approximation is valid we find that
momentum coupling indeed causes an increase of the rate
which is even greater than the classical enhancement. This
is due to the thinning of the parabolic barrier. At low
temperatures, below the crossover temperature, momentum
coupling leads even to an exponential increase in the
tunneling rate.

\subsubsection{Above the crossover temperature}\label{sec:3b1}

In the limit of a parabolic barrier, we follow the same
reasoning as for spatial diffusion, as given in
Ref.~\cite{pollak86}. In the
absence of momentum coupling, the thermal fraction through the barrier is%
\begin{equation}
F_{0}^{Q}=2^{-N}\frac{\hbar \Omega ^{\ddag }}{2}\sin \left(
\frac{\hbar \beta \Omega ^{\ddag }}{2}\right)
^{-1}\prod\limits_{j=1}^{N}\sinh \left( \frac{\hbar \beta \omega
_{j}}{2}\right) ^{-1}  \label{3.8}
\end{equation}%
while the VTST tunneling fraction is given by the same expression, except
one must replace the bare frequencies everywhere with the normal mode
frequencies. One thus has that the quantum transmission factor is
\begin{equation}
\frac{F_{VTST}^{Q}}{F_{0}^{Q}}=\frac{\lambda ^{\ddag }}{\Omega ^{\ddag }}%
\frac{\sin \left( \frac{\hbar \beta \Omega ^{\ddag }}{2}\right) }{\sin
\left( \frac{\hbar \beta \lambda ^{\ddag }}{2}\right) }\prod%
\limits_{j=1}^{N}\frac{\sinh \left( \frac{\hbar \beta \omega _{j}}{2}\right)
}{\sinh \left( \frac{\hbar \beta \lambda _{j}}{2}\right) }.  \label{3.9}
\end{equation}%
However we know that the barrier frequency $\lambda^{\ddag}\geq
\Omega^{\ddag}$. The denominator with the $\sin$ function will
diverge at the temperature $\hbar\beta_c\lambda^{\ddag}=2\pi$, that
is the crossover temperature will be {\it greater} than the
crossover temperature in the absence of coupling. In contrast to
spatial coupling, which reduces the crossover temperature, momentum
coupling increases it. At the crossover temperature, the quantum
transmission factor will diverge, implying that the quantum
transmission factor is indeed {\it larger} than the classical one.
As noted, for the parabolic barrier, momentum coupling enhances
tunneling.

The result for the tunneling fraction has to be
transformed so that it may be expressed in the continuum
limit. For this purpose one uses the infinite product
representation of the $\sin $ and $\sinh $ functions (as
detailed in Ref.~\cite{abramowitz72})
as well as the identity:%
\begin{equation}
\det \left( \mathbf{T^{\prime \prime
}}+s^{2}\mathbf{I}\right) =\left( -\lambda ^{\ddag
2}+s^{2}\right) \prod\limits_{j=1}^{N}\left( \lambda
_{j}^{2}+s^{2}\right)
=\left( -\frac{\Omega ^{\ddag 2}}{M}+s^{2}+\Omega ^{\ddag 2}\frac{\hat{\eta}%
_{P}(s)}{s}\right) \prod\limits_{j=1}^{N}\left( \omega
_{j}^{2}+s^{2}\right)   \label{3.10}
\end{equation}%
where as before $\mathbf{T^{\prime \prime }}$ represents
the matrix of second derivatives of the kinetic energy with
respect to the (frequency scaled) momenta. The last
equality on the r.h.s is obtained by carrying out
explicitly the evaluation
of the determinant, using the frequency scaled momenta, as given in (\ref{2.15}) and (\ref{2.16}).
One then finally finds that%
\begin{equation}
\frac{F_{VTST}^{Q}}{F_{0}^{Q}}=\frac{\lambda ^{\ddag }}{\Omega ^{\ddag }}%
\prod\limits_{k=1}^{\infty }\frac{\nu_{k}^{2}-\Omega ^{\ddag 2}}{%
\nu_{k}^{2}-\frac{\Omega ^{\ddag 2}}{M}+\Omega ^{\ddag 2}\frac{\hat{%
\eta}_{P}(\nu_{k})}{\nu_{k}}}\equiv \frac{\lambda ^{\ddag }}{%
\Omega ^{\ddag }}\, \Xi _{P}  \label{3.11}
\end{equation}%
where $\nu_{k}=\frac{2k\pi }{\hbar \beta }$ are the Matsubara
frequencies and $\Xi _{P}$ is the Wolynes factor \cite{wolynes81}
associated with momentum coupling.

\begin{figure}
\begin{center}
\vspace*{0.1cm} \epsfig{file=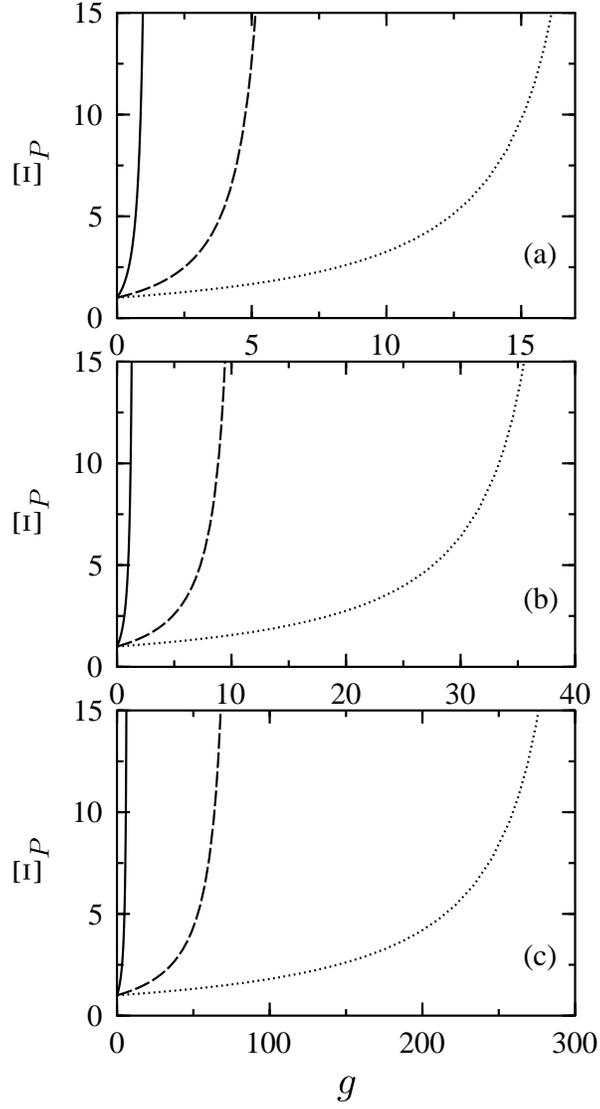, height=15cm}
\caption[]{\label{fig_3}Macroscopic enhancement of the
quantum tunneling rate above crossover due to momentum
coupling to a harmonic bath. Panels (a)--(c) correspond to
the cutoff frequencies $10, 4$ and $0.5$, respectively. In
each panel we plot the dependence of the Wolynes
enhancement factor as a function of the coupling strength
for three different temperature values. Defining
$\theta=\hbar\beta\Omega^{\ddag}$, the dotted, dashed and
solid lines correspond to $\theta=0.5,1,3$ respectively.
Note the change in scale for the coupling strength as one
goes from panel (a) to panel (c). For low cutoff
frequencies, the enhancement is weakened.}
\end{center}
\end{figure}

For the specific spectral density given in (\ref{spectral})
one finds that the Wolynes
factor is:%
\begin{equation}
\Xi _{P}=\prod\limits_{k=1}^{\infty
}\frac{\nu_{k}^{2}-\Omega ^{\ddag 2}}{\nu_{k}^{2}-\Omega
^{\ddag 2}\left( 1+\gamma \nu_{k}\arctan \left(
\frac{\omega _{c}}{\nu_{k}}\right) \right) } \label{3.12}
\end{equation}
showing clearly that the quantum enhancement of the rate
due to momentum coupling is \textit{greater}{\huge \ }than
the classical enhancement, since in the classical limit
the Wolynes factor tends to unity. In Fig.~\ref{fig_3} we
plot this Wolynes factor as a function of the reduced
momentum coupling coefficient $g=\gamma\Omega$ and reduced
cutoff frequency $w_{c}=\omega_c/\Omega$.

\subsubsection{Tunneling through an anharmonic barrier below the crossover temperature}\label{sec:3b2}

For tunneling through an anharmonic barrier potential
$V(Q)$ with a barrier top located at $Q=Q_b$ the escape
rate $\Gamma$ is most conveniently calculated from the
imaginary part of the free energy and thus from the
imaginary part of the partition function of the unstable
system \cite{caldeira83,weiss95}. In the path integral
representation one has for the reduced system
\begin{equation}
Z= {\rm Tr}\{{\rm e}^{-\beta H}\}=\oint {\cal D}[Q]\ {\rm
e}^{-S_{\rm eff}[Q]/\hbar}\, ,
\end{equation}
where one sums over all periodic paths running in the
imaginary time interval $\hbar\beta$ through the inverted
barrier potential $-V(q)$. Here for momentum friction the
effective action is found to read
\begin{equation}
S_{\rm eff}[Q]=\int_0^{\hbar\beta}d\tau \left\{
\frac{1}{2}\dot{Q}(\tau)^2+V(Q)+\frac{1}{2}\int_0^{\hbar\beta}d\sigma
\dot{Q}(\tau)\,
\tilde{M}(\tau-\sigma)\,\dot{Q}(\sigma)\right\}
\label{effaction}
\end{equation}
with the kernel given by
\begin{eqnarray}
\tilde{M}(\tau)&=&\frac{1}{\hbar\beta}\sum_n{\rm e}^{i\nu_n
\tau}\left(\frac{1}{1+\sum_j
d_j^2\frac{\nu_n^2}{\nu_n^2+\omega_j^2}}-1\right)\nonumber\\
&=&-\frac{1}{\hbar\beta}\sum_n{\rm e}^{i\nu_n \tau}\,
\frac{\sum_j
d_j^2\frac{\nu_n^2}{\nu_n^2+\omega_j^2}}{1+\sum_j
d_j^2\frac{\nu_n^2}{\nu_n^2+\omega_j^2}}\nonumber\\
&=& -\frac{1}{\hbar\beta} \sum_n\, m_n\, {\rm e}^{i\nu_n
\tau}\, .
\end{eqnarray}
For high temperatures when $\nu_n\to \infty$ one has
$\tilde{M}(\tau)=:\delta(\tau): \, M \sum_j d_j^2$ with
$M=1/(1+\sum_j d_j^2)$. Combining the friction term with
the bare kinetic term then leads to an effective kinetic
term of the form $M \dot{Q}^2/2$ so that friction appears
simply as an effective mass. Accordingly, the thermal
activation factor $\Gamma\propto \exp[-\beta V(Q_b)]$ is
independent of friction since it is determined by the
constant path at the barrier top $Q=Q_b$, while
dissipation influences the prefactor as specified above.
The same is true for lower temperatures above the crossover
when quantum fluctuations in the rate prefactor lead  to
an even stronger increase of the escape rate as shown in
the previous Section.

Below the crossover temperature the contribution of the bounce orbit
$Q_B$ dominates the imaginary part of the partition function and the
exponential factor in the rate contains its action $S_B=S[Q_B]$,
i.e.\ $\Gamma\propto \exp(-S_B/\hbar)$. Now, let us consider weak
friction. In this case we write for the bounce path $Q_B=Q_0+\delta
Q$, where $Q_0$ is the bounce path in absence of dissipation, which
obeys $\ddot{Q}_0-V'(Q_0)=0$. Upon inserting $Q_B$ into the
effective action (\ref{effaction}) one finds that to lowest order in
the friction one has $S_B=S[Q_0]+\Delta S_0$ with
\begin{equation}
\Delta S_0=\frac{1}{2}\int_0^{\hbar\beta}d\tau
\int_0^{\hbar\beta}d\sigma \, \dot{Q}_0(\tau)\,
\tilde{M}(\tau-\sigma)\, \dot{Q}_0(\sigma).
\end{equation}
This correction can be easily expressed in Fourier space
by using $Q_0=(1/\hbar\beta)\sum_n Q_n^{(0)}
\exp(i\nu_n\tau)$, where one may choose the phase of the
bounce such that $Q_n^{(0)}=Q_{-n}^{(0)}={Q_n^{(0)}}^*$.
This way, we find with $m_n=m_{-n}$ and $\nu_n=-\nu_{-n}$
that
\begin{equation}
\Delta S_0=-\frac{1}{2\hbar\beta}\sum_n\, |Q_n^{(0)}|^2\,
\nu_n^2\, m_n\, .
\end{equation}
Apparently, $\Delta S_0<0$ so that $S_B<S_0$ meaning that
the probability for quantum tunneling is exponentially {\em
enhanced} due to momentum friction. Physically, since the
effective mass of the combined kinetic terms
$\tilde{M}(\tau)+:\delta(\tau):$ has Fourier components
that are always smaller than 1, the particle's kinetic
energy becomes smaller relative to its potential energy.
Thus, for a dynamical orbit like the bounce path the action
decreases. In contrast, spatial friction leads always to a
rate suppression since effectively it provides an
additional contribution to the potential energy. For
stronger coupling to the heat bath the bounce orbit can
easily be calculated numerically along the lines described
in \cite{grabert87}.

\section{Discussion}\label{sec:4}

In this paper we analyzed the influence of a harmonic heat bath
coupled to a system with potential energy via a bilinear interaction
between the system's momentum and the individual momenta of the bath
degrees of freedom. For a harmonic oscillator and momentum
dissipation we find as also noted qualitatively by Cuccoli et al
\cite{cuccoli01} that the coupling to the bath {\it increases} the
delocalization of the position of the oscillator. The thermal
variance {\it increases} with increasing coupling, exactly the
opposite of the behavior found for spatial coupling.

For escape over a barrier, in the high temperature regime, where
thermal activation prevails, momentum friction leads to an increase
of the flux across the barrier. At somewhat lower temperatures
quantum fluctuations come into play and enhance the flux even more.
Above the crossover temperature between tunneling and thermal
activation, friction appears only in the prefactor of the rate
expression. Below the crossover temperature, where quantum tunneling
dominates, it reduces the action of the bounce path and thus
exponentially increases the decay rate. In contrast to the case of
spatial friction, the crossover temperature increases with
increasing friction so that for pure momentum coupling quantum
effects would be observable even in the high temperature domain.

In this paper we considered the case of pure momentum coupling. Any
realistic system will be influenced by both, momentum {\em and}
spatial coupling. The crucial issue is then to what extent the
latter one suppresses the impact of the former one. Specific systems
to be studied in the future in this respect include molecular
compounds, mesoscopic islands coupled to fluctuating charges and the
transport of charged particles moving under the influence of random
magnetic fields \cite{fermi49,sturrock66}.

\acknowledgments

We thank Prof. Y. Makhnovskii for his comments on the manuscript. We
also thank the anonymous referees for their comments and especially
for bringing to our attention Refs. \cite{leggett84,cuccoli01}. This
work has been supported by grants from the German Israel Foundation
for Basic Research and the Israel Science Foundation.

\end{document}